\documentclass[sigconf]{acmart}

\usepackage{enumitem}

\AtBeginDocument{%
  \providecommand\BibTeX{{%
    \normalfont B\kern-0.5em{\scshape i\kern-0.25em b}\kern-0.8em\TeX}}}

\copyrightyear{2022}
\acmYear{2022}
\setcopyright{acmlicensed}\acmConference[EASE 2022]{The International Conference on Evaluation and Assessment in Software Engineering 2022}{June 13--15, 2022}{Gothenburg, Sweden}
\acmBooktitle{The International Conference on Evaluation and Assessment in Software Engineering 2022 (EASE 2022), June 13--15, 2022, Gothenburg, Sweden}
\acmPrice{15.00}
\acmDOI{10.1145/3530019.3534082} \acmISBN{978-1-4503-9613-4/22/06}

\begin{document}
\sloppy

\title[Understanding the Characteristics of Visual Contents in Open Source Issue Discussions]{Understanding the Characteristics of Visual Contents in Open Source Issue Discussions: A Case Study of Jupyter Notebook}

\author{Vishakha Agrawal}
\authornote{Both authors contributed equally to this research.}
\email{vishakha.agrawal09@gmail.com}
\affiliation{%
  \institution{Dayananda Sagar College of Engineering}
  \city{Bangalore}
  \country{India}
}

\author{Yong-Han Lin}
\authornotemark[1]
\email{alberta.cs07@nycu.edu.tw}
\affiliation{%
  \institution{National Yang Ming Chiao Tung University}
  \city{Hsinchu}
  \country{Taiwan}
}

\author{Jinghui Cheng}
\email{jinghui.cheng@polymtl.ca}
\affiliation{%
  \institution{Polytechnique Montreal}
  \city{Montreal}
  \state{QC}
  \country{Canada}
}

\begin{abstract}
    Most issue tracking systems for open source software (OSS) development include features for community members to embed visual contents, such as images and videos, to enhance the discussion. Although playing an important role, there is little knowledge on the characteristics of the visual contents to support their use. To address this gap, we conducted an empirical study on the Jupyter Notebook project. We found that more than a quarter of the issues in the Jupyter Notebook project included visual contents. Additionally, issues that had visual contents in the comments but not in the issue posts tended to have a longer discussion and to involve a larger number of discussants. In our qualitative analysis, we identified eight types of visual contents in our sample, with about 60\% including screenshots or mockups of the main product. We also found that visual contents served diverse purposes, touching both problem and solution spaces of the issues. Our effort serves as an important step towards a comprehensive understanding of the characteristics of visual contents in OSS issue discussions. Our results provided several design implications for issue tracking systems to better facilitate the use of visual contents.
\end{abstract}

\ccsdesc[500]{Software and its engineering~Open source model}
\ccsdesc[500]{Human-centered computing~Empirical studies in collaborative and social computing}


\keywords{Visual contents, open source software, issue discussions}

\maketitle

\section{Introduction}
Many open source software (OSS) projects rely on issue tracking systems (e.g., GitHub Issues) to allow diverse community members to collaboratively contribute to the OSS project~\cite{heck2013analysis}. The issue reports and the corresponding discussions constitute a rich bank of resources that includes abundant information and serves various roles in the OSS projects, such as bug tracking, enhancements and new features discussion, task management, question and answer, and community engagement, to just name a few~\cite{arya2019analysis}. Most issue tracking systems do not only allow textual discussions, but they also include features for OSS community members to embed visual contents, such as images and videos, to have an enhanced discussion experience.

These visual contents play an important role in issue discussions. Previous research has identified that software developers considered visual contents such as screenshots as important elements in bug reports~\cite{Bettenburg2008what, Breu2010}. In usability and GUI issues in particular, visual contents play a crucial role in communicating the cause of the problem, the proposed solution, the issue context, and the observed results~\cite{Yusop2016}. Our preliminary exploration of the visual contents also indicated that they are commonly seen in non-GUI issues to show valuable elements such as code snippets or error messages.

Although the visual contents are widely used and largely valued in issue discussions, there is little research focused on a comprehensive understanding of their characteristics to support their use. With the increasing diversity of OSS community members that start to involve more end users, UX designers, and casual contributors, this knowledge becomes imperative. A clear understanding of the use of visual contents can provide important information to support the diverse OSS community members in engaging in more efficient discussions and to inform the creation of new issue tracking tools (including automated tools) that can better leverage this type of important communication media. In this paper, we directly target these gaps by posing the following research questions.

\begin{enumerate}[noitemsep, leftmargin=*, itemindent=0pt, label={\textbf{RQ{{\arabic*}}}:}]
    \item How frequently do people use visual contents in issue discussions? 
    \item How is the usage of visual contents associated with issue characteristics, such as issue status, discussion lengths and discussant participation?
    \item What are the common types of visual contents posted in the issue tracking system?
    \item What are the common purposes for posting visual contents?
\end{enumerate}

To answer our research questions, we targeted the Jupyter Notebook project as a case study. This project is a web-based data science notebook that is under active development and involves a large and diverse community. Thus it has the potential of including diverse types of visual contents in its issue discussions. For RQ1 and RQ2, we conducted a statistical analysis on the entire dataset of 4210 issues in this project. Our results indicated that a large quantity (25.4\%) of the discussion threads included one or more visual contents. We also found that the issues which included visual contents in the comments but not in the issue post were associated with a larger number of comments, involving a larger amount of discussants. To answer RQ3 and RQ4, we conducted a qualitative content analysis on a sample of 50 issues that included 124 visual contents. We identified eight types of visual contents used in those issues, with a considerable amount (36.2\%) not related to the user interface of the product. We also found that the issue discussants used the visual contents to support the discussion on both the problem and solution spaces of the issues, touching across various problem-solving stages. Overall, our study serves as an important step towards comprehensive knowledge about the use of visual contents and provided useful implications and suggestions to researchers and practitioners who aim to investigate and enrich the open source issue discussion experience.
\section{Related Work}
Previous studies have established that Issue Tracking Systems (ITSs) supports various software engineering activities such as requirements identification~\cite{Heck2017}, feature request detection~\cite{Merten2016}, bug triaging~\cite{Xia2017}, design rationale retrieval~\cite{Viviani2018}, and software traceability~\cite{Nicholson2020}, to name a few. Many modern ITSs such as GitHub Issues provide extended features for OSS community members to engage in discussions of diverse topics, making ITSs to embed a rich body of information~\cite{arya2019analysis}.

Many recent studies have focused on how OSS community members interact in the issue discussion threads. For example, Rath and M\"{a}der~\cite{Rath2020} identified three patterns of issue discussions: (1) monolog, (2) feedback, and (3) collaboration. Sanei et al.~\cite{Sanei2021} found that affective states expressed in OSS issue discussions have a complex association with many issue characteristics. Nurwidyantoro et al.~\cite{Nurwidyantoro2022} identified that issue discussions do not only express \textit{system value themes}, but they also are a rich source of \textit{human value themes} such as dignity, inclusiveness, and privacy. Researchers have also investigated negative social interactions, such as incivility~\cite{Ferreira2021Incivility} and toxicity~\cite{Miller2022Toxicity}, that happened in ITSs and related software engineering artifacts. To help address the complexity of discussions in ITSs, Wang et al.~\cite{Wang2020ArguLens} proposed a technique leveraging an argumentation model to organize the different perspectives and their supporting arguments made by diverse OSS community members. Our study is built on this rich body of related literature on ITSs and focus on the roles of visual contents in the issues discussions.

While there is little work directly targeting visual contents in ITSs, previous studies that focused on understanding general usage patterns of ITSs have established that visual contents play an important role. Intuitively, the use of visual contents supports the discussion of usability and user interface issues~\cite{Folstad2012, Yusop2016}. However, studies also revealed the importance of visual contents in discussions beyond usability issues. For example through a survey study, Bettenburg et al.~\cite{Bettenburg2008what} identified that although developers often need visual contents such as screenshots to understand the issue, bug reporters sometimes overlooked the importance of such contents. Related, Breu et al.~\cite{Breu2010} found that information such as examples, program output, and screenshots are often missing in the bug reports and frequently requested by the developers. Similarly, Davies and Roper~\cite{Davies2014} identified that visual contents such as screenshots were often not included in the original bug report but provided as an attachment in a later comment. Although screenshots provided useful information, the authors expressed concerns about utilizing such information due to the challenges involved in analyzing these visual contents~\cite{Davies2014}. 
A more optimistic view is discussed by Nayebi in a recent vision paper~\cite{Nayebi2020}. The author argued that, with the trend of using visual contents in software development and the support of machine learning techniques, extracting useful information from these contents would become increasingly relevant. A few concrete attempts were made in this direction. For example, Wang et al.~\cite{WANG2019} proposed a technique that combines the screenshot image features with textual features to detect duplicated bug reports.

In sum, previous studies indicated the importance of visual contents in ITSs and the potential of leveraging information embedded in such contents. However, there is little research focused on a comprehensive understanding of their characteristics, while this knowledge is essential to inform tools and techniques to facilitate the use of visual contents. Our study fills this exact gap.
\section{Methods}

\subsection{Project selection and issues sampling}
We focused our study on the repository of Jupyter Notebook (https://github.com/jupyter/notebook), which is a web-based data science notebook project hosted on GitHub. We chose this project mainly because of two reasons. First, the project is under active development and attracting a large community involving diverse contributors. Second, all authors of this paper are active users of this project, making comprehensive and accurate analysis possible.

Data collection was done in May 2021. In order to collect the issues that contain visual contents in the Jupyter Notebook project for analysis, we downloaded all 4210 open and closed issues, along with comments made to the issues, using the GitHub REST API. We then used the following two criteria to identify visual contents in the body of issue reports and comments:
(1) the textual body needs to include a link whose URL contains one of the following domain names: \textit{user-images.githubusercontent}, \textit{cloud.githubusercontent}, \textit{camo.githubusercontent} and \textit{raw.githubusercontent}; and (2) the link URL contains one of the following file extensions: .gif, .png, .jpg, .jpeg, .mp4, and .mov. These criteria helped us to identify both image and video contents and identify contents presented in both Markdown syntax and HTML syntax. These criteria helped us to identify 1071 issues in the Jupyter Notebook project that contained at least one visual content in their discussion threads.

\subsection{Quantitative Analysis Methods}
To answer RQ1 and RQ2, we conducted a statistical analysis on the entire dataset of 4210 issues. We treated the existence of a visual content as an independent variable and considered three levels: (1) no visual contents, (2) with visual contents in the issue post, and (3) with visual contents not in the issue post, but in at least one comment. We then considered three dependent variables, namely (1) the state of the issue (i.e., open or closed), (2) the discussion length indicated by the number of comments, and (3) the number of unique discussants participated. We conducted a Pearson's chi-squared test (for the first dependent variable) and one-way ANOVA tests (for the other two dependent variables) to compare the differences among the three independent groups. If a significant difference was identified, pairwise posthoc analysis was conducted to understand the groups that contributed to the difference.

\subsection{Qualitative Content Analysis Methods}
To answer RQ3 and RQ4, we conducted a qualitative content analysis~\cite{Vaismoradi2013} on a random sample of 50 issues among all the 1071 issues that contained at least one visual content. The sampled issues contained a total of 124 visual contents with a median of 2 visual contents per issue (range from 1 to 14). First, two annotators independently performed inductive coding on: (1) the \textit{type} of each visual content and (2) the \textit{purpose} served by each visual content in the discussion. To identify the purpose, we analyzed the comment that included a visual content and the comments before it to capture the discussion context. Once the individual coding was completed, all three authors met and exercised an affinity diagramming activity to group the individual codes into higher-level categories. These categories were then recorded in a codebook detailing criteria for identifying each category. After the meetings, two original annotators used the codebook to code all the visual contents again and calculated inter-rater reliability using Cohen's kappa. Among all the codes included in the codebook, the average kappa coefficient is 0.67 ($SD=0.21$), indicating a substantial agreement~\cite{viera2005}. The two annotators then discussed and resolved their disagreements, resulting in the final coding of the dataset.

\section{Results}
\label{sec:results}


\subsection{RQ1: Frequencies of Visual Contents Use}
\begin{figure}[t]
    \centering
    \includegraphics[width=0.75\columnwidth]{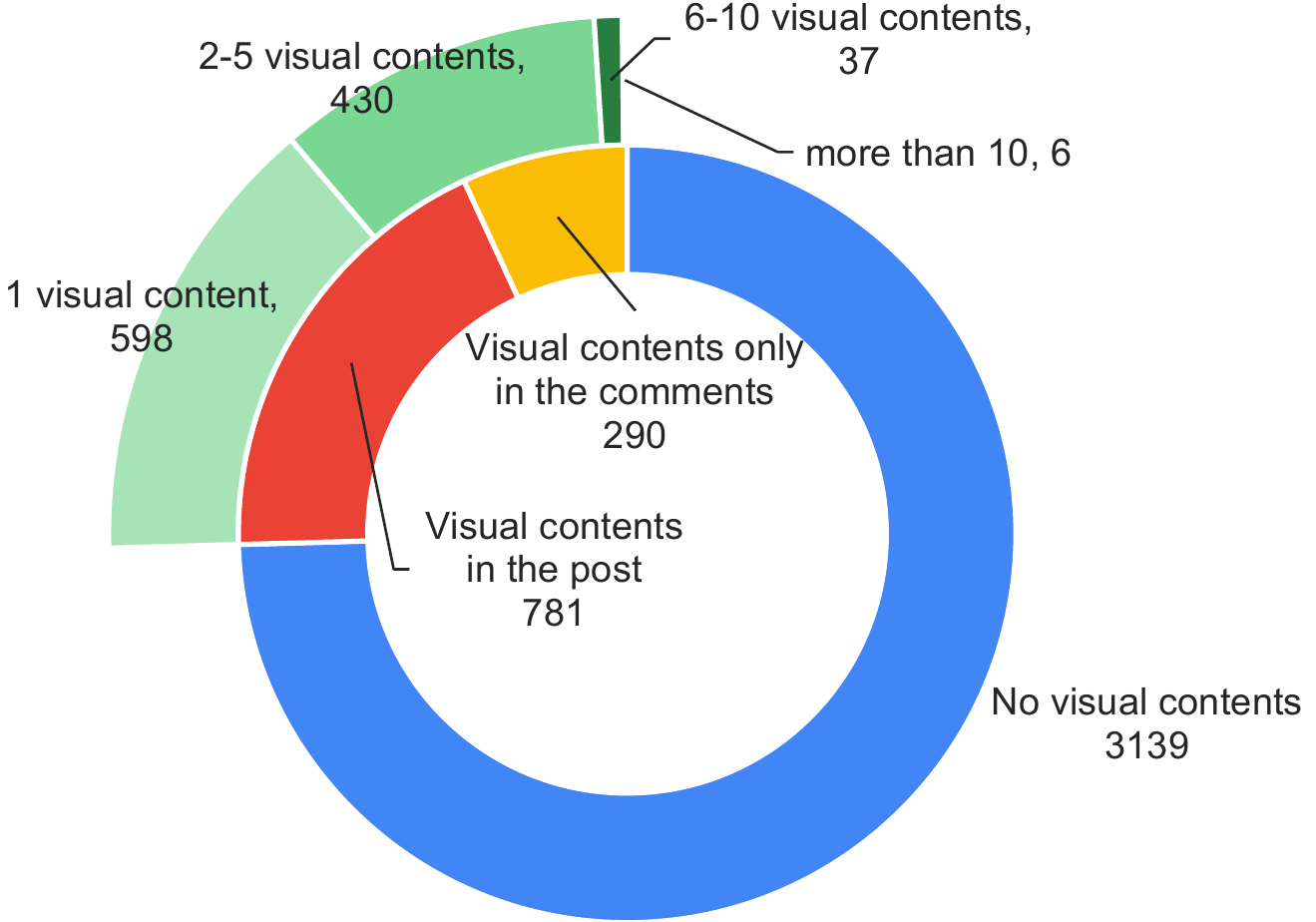}
    \caption{Frequency of issues that included visual contents}
    \vspace{-12pt}
    \label{fig:frequency}
\end{figure}

Among the 4210 issues in the Jupyter Notebook project, there were 1071 issues (or 25.4\%) that contained at least one visual content in either the issue report or one of the comments. Among the issues that contained a visual content, 781 issues contained visual contents in the issue post and 290 issues contained visual contents only in the comments. Figure~\ref{fig:frequency} summarizes the results.

In total, we found 2047 visual content instances among these issues, including 1159 in issue reports and 888 in comments. Most of the issues ($N=598$) contained only one visual content, while 430 issues contained two to five, 37 issues contained six to ten, and six issues contained more than ten visual contents. The maximum number of visual contents in an issue was 27 (Issue\#5692), followed by 17 (Issue\#5364). An unpaired t-test did not find a significant difference ($t=0.87$, $p=0.39$) on the number of visual contents per issue between the issues with visual contents in the post ($mean=1.94$) and those only in the comments ($mean=1.83$).

\subsection{RQ2: Association with Issue Characteristics}
Recall that to answer RQ2, we considered three independent groups: (1) issues with no visual contents ($N=3139$), (2) issues with visual contents in the issue post ($N=781$), and (3) issues with visual contents only in the comments ($N=290$).

\textit{State of Issues.}
Among all 4210 issues, there were 2226 (52.9\%) closed issues by the time of our data collection. Among issues with no visual contents, 1655 (52.7\%) were closed; the numbers of closed issues for the other two independent groups were 399 (51.1\%) and 172 (59.3\%), respectively. A Pearson's chi-square test indicated that this difference is not statistically significant ($chi^2=5.85$, $p=0.054$).

\textit{Length of Discussion.}
Through a one-way ANOVA test, we found a statistically significant difference in the length of discussion among the three groups of issues ($F=256.07$, $p<0.0001$). A Tukey HSD post hoc test revealed that the differences between each pair of the independent groups are significant. Specifically, issues with visual contents only in the comments ($p<0.01$) involved significantly longer discussions ($mean=15.50$, $SD=17.87$) than those with visual contents in the issue post ($mean=5.59$, $SD=8.39$), which in turn was significantly longer ($p=0.012$) than those without visual contents ($mean=4.70$, $SD=5.83$).



\textit{Discussants Participation.}
The one-way ANOVA test also revealed a statistically significant difference in the number of unique discussants participated among the three groups of issues ($F=203.13$, $p<0.0001$). Using a Tukey HSD post hoc test, we found that the number of unique discussants in issues with visual contents only in the comments ($mean=8.14$, $SD=11.25$) is significantly larger ($p<0.01$) than those with visual contents in the issue post ($mean=3.17$, $SD=4.18$) and those without visual contents ($mean=2.81$, $SD=2.99$); the difference between the latter two groups was not significant ($p=0.10$).



\subsection{RQ3: Types of Visual Contents}
\begin{figure}[t]
    \centering
    \includegraphics[width=0.8\columnwidth]{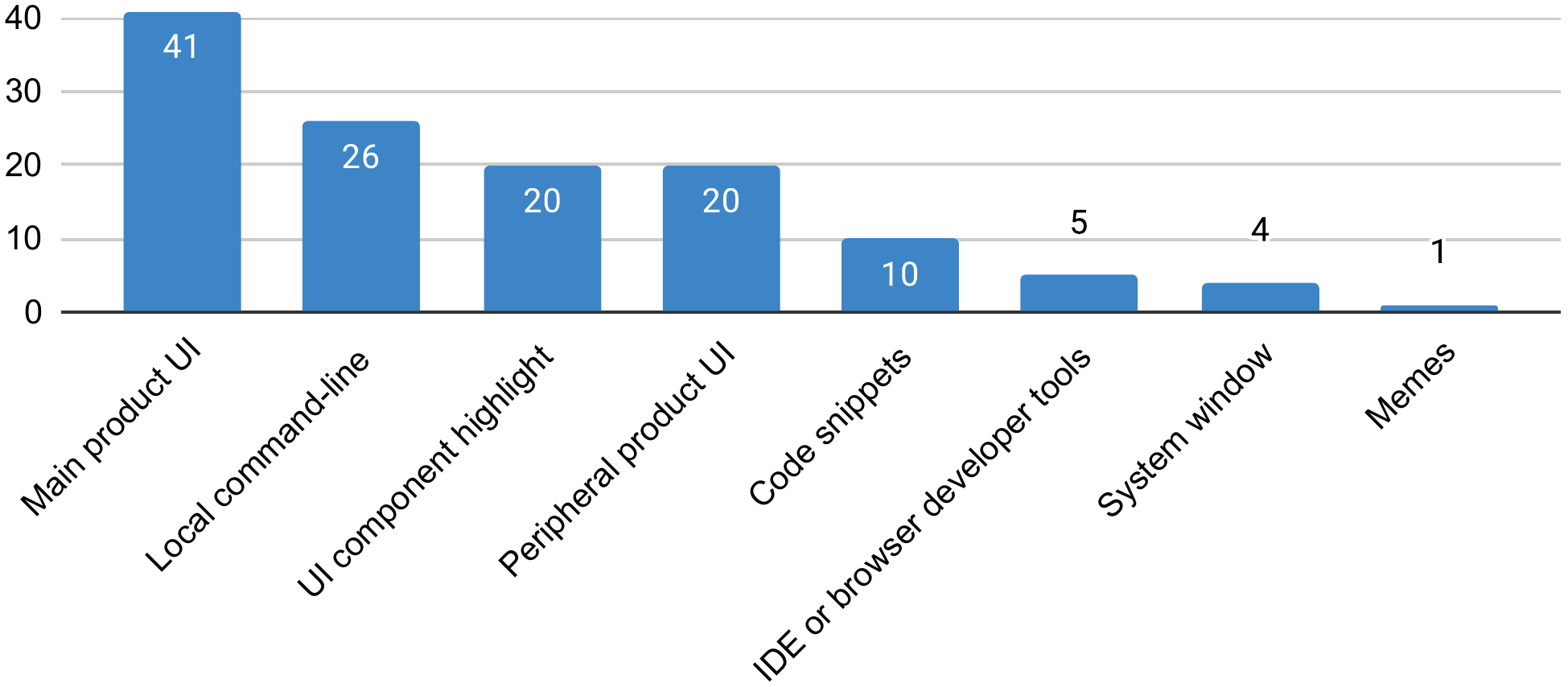}
    \caption{Frequency of types of visual contents}
    \vspace{-12pt}
    \label{fig:content-type-frequency}
\end{figure}

In our qualitative analysis, we identified eight types of visual contents used in the issue discussions. Each visual content can have multiple parts that belong to different types. Figure~\ref{fig:content-type-frequency} summarizes the frequency of these content types, which we describe below.

\emph{Main product UI.} In 41 cases, the discussants included a screenshot or animation to show the main UI of the project. In the Jupyter Notebook project, this means the working area of the notebook and includes cells, command palette, cell outputs, the error console, etc. For example, the first visual content in Issue\#252 shows an animation of the execution of the code in the Jupyter environment.

\emph{Local command-line terminal/console.} In 26 cases, the discussants showed command-line terminals or consoles of their local operating system, usually for illustrating local configurations and/or installation processes. For example, in Issue\#172, a discussant used screenshots of Windows PowerShell to demonstrate the commands for setting up a Ubuntu environment in Windows 10.

\emph{UI component highlight.} In 20 cases, the discussants included screenshots or animations to show a highlighted UI component (e.g., toolbar, menu bar, status bar, icon, pop-up window, etc.), either in an existing UI or as a mockup. For example, in Issue\#533 MM, discussants used highlighted UI components to illustrate their feedback to the UI design before a release.

\emph{Peripheral product UI.} In 20 cases, the discussants posted pages or sections that were not the main working area of the product. These included the file management page, the home page, the about page, etc. For example, in Issue\#244, a discussant used a screenshot of the About page of Jupyter to show how to find package versions that the current instance of the Jupyter server is running on.

\emph{Code snippets.} In 10 cases, the discussants used images, instead of formatted texts, to show code snippets. 
For example, a discussant in Issue\#5054 used images to show code snippets before and after a change that fixed their issue.

\emph{IDE or browser developer tools.} In five cases, the discussants showed the screenshots of IDE (for the back-end) or browser developer tools (for the front-end) to illustrate development issues. For example, a discussant in Issue\#469 used a screenshot of a web inspector to show an issue in the front-end code.

\emph{System window.} In four cases, the discussant showed a GUI window of their local operating system, such as the setting window, the task manager, and the file manager. For example, a discussant in Issue\#172 used screenshots of the Windows settings page to show how to turn on developer mode.

\emph{Memes.} In one case (Issue\#254), a discussant posted a GIF of cats.

\subsection{RQ4: Purposes of Posting Visual Contents}
\begin{figure}[t]
    \centering
    \includegraphics[width=0.85\columnwidth]{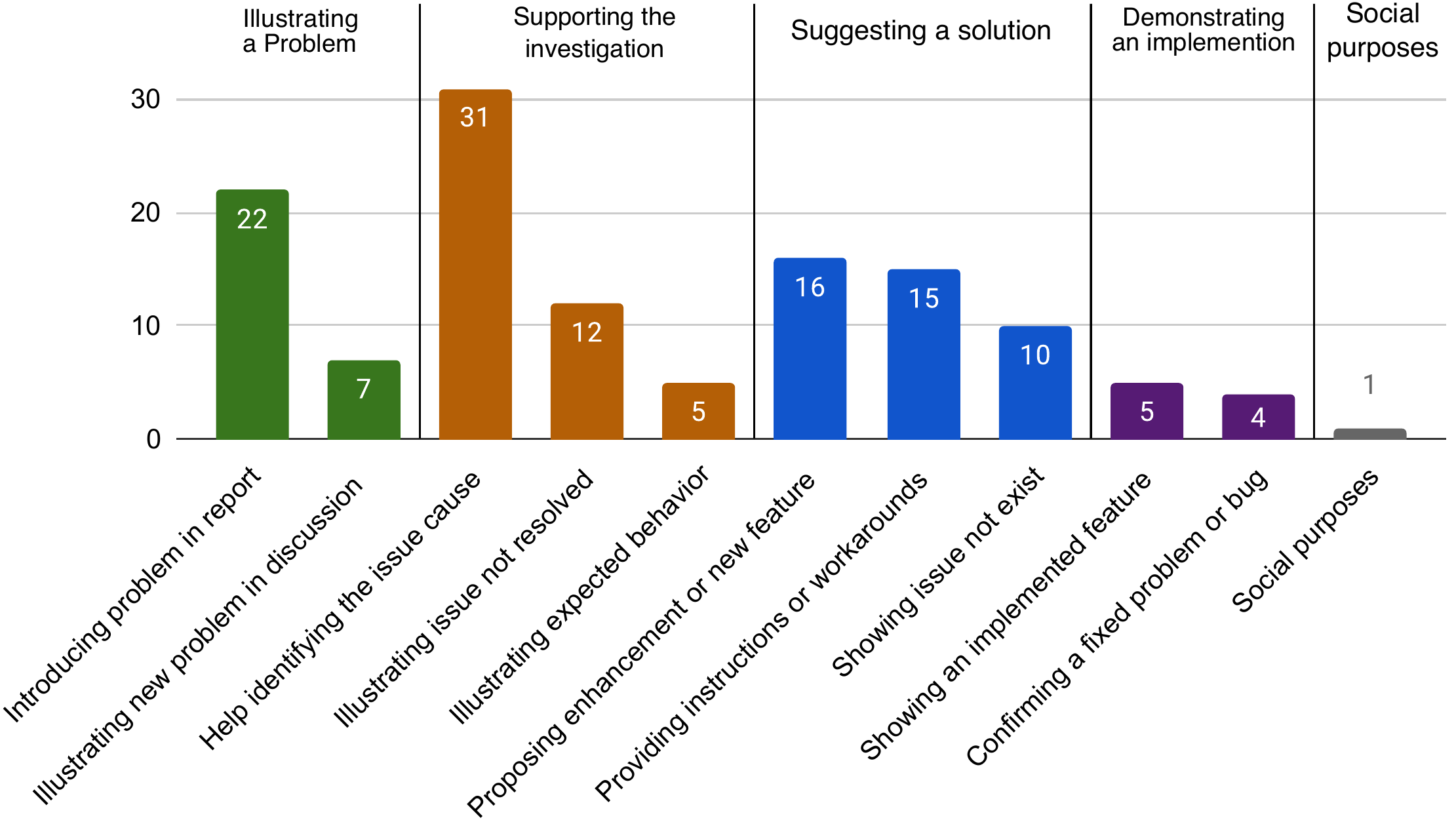}
    \caption{Frequency of purposes of posting visual contents}
    \vspace{-12pt}
    \label{fig:purpose-frequency}
\end{figure}

We identified five general purposes why a discussant used a visual content in the issue discussion. In each of the general purposes, we also identified specific themes that we describe below. Figure~\ref{fig:purpose-frequency} illustrate the frequencies of the general and the specific purposes that we identified in our analysis.

\subsubsection{Illustrating a Problem}
In our sample, we identified 29 visual contents that were used for demonstrating a problem related to the software product. Within this general purpose, discussants focused on the following two specific purposes when using a visual content.

\emph{Introducing the problem in the issue report.} In 22 cases, discussants used visual contents to illustrate the problem introduced in the issue post. For example, in Issue\#110 the issue reporter shows a UI problem that the ``...'' button is not properly aligned. Sometimes, discussants used visual contents to compare two versions of the system to show that a recent change is not desirable. For example, a discussant reported an issue (Issue\#5629) and used screenshots to illustrate the difference before and after a code commit.

\emph{Illustrating a new problem encountered during discussion.} In seven cases, the discussant used visual contents to illustrate new problems found during the issue discussion and posted them as issue comments. These new problems may or may not be related to the original issue. For example, in Issue\#3506, a discussant used screenshots to illustrate a problem they faced in a new version of the system, which is entirely different from the original issue report.



\subsubsection{Supporting the investigation of a problem}
We found 48 visual contents used for the general purpose of supporting the investigation of a problem, with the specific purposes we describe below.

\emph{Providing information for identifying the issue cause.} In 31 cases, the discussant used visual contents to show additional information about the issue  (e.g., error message, additional symptoms, behavior in another environment, etc.) to support the identification of its cause. For example, in Issue\#430, a discussant posted a screenshot of an error message after following an instruction step (the 7th comment) in order to support the discussion about the issue cause.

\emph{Illustrating that the issue is not resolved.} We found that in 12 cases, the discussant used visual contents to illustrate that they are unable to follow the instructions to resolve an issue or to support that the bug is reproduced in a certain environment. For example, in the 6th comment in Issue\#290, a discussant used a screenshot to show that they followed an instruction step provided by another discussant but the problem still existed. 

\emph{Illustrating expected behavior.} In five cases, the discussant used visual contents (usually a mockup) to illustrate the expected behavior of the system without a bug. Sometimes, the discussants used two visual contents to compare the current behavior and the expected behavior. For example, in the 5th comment of Issue\#533, the discussant shows that a UI element disappeared in a new version.

\subsubsection{Suggesting a solution}
We found that in 41 cases, discussants used visual contents to suggest a solution to the issue. This includes the following three specific purposes.

\emph{Proposing enhancement or new feature ideas.} In 16 cases, discussants used images or animations to illustrate enhancements or new feature ideas; the end goal was to request the feature or to ask for feedback on their ideas. The ideas can be shown by annotating a screenshot or by creating a mockup. For example, in the 12th comment of Issue\#558, a discussant made a mockup for an enhancement to a Find and Replace feature. Sometimes, the feature idea was inspired by a competitor project and the screenshots of that project were used to illustrate the idea. 

\emph{Providing instructions or workarounds to a usage issue.} In 15 cases, discussants included visual contents to show how to perform a certain task on the system, either as instructions or workarounds to help the other participants resolve the issue. These contents were also used to provide additional details about a previously provided resolution or a newly implemented feature. For example, in the 12th comment of Issue\#172, the discussant posted instruction steps, illustrated with screenshots, to help resolve the issue about running a Jupyter terminal inside Windows.

\emph{Showing that the issue does not exist.} In 10 cases, visual contents were used to show that the reported bug was actually a feature, the bug was not reproducible, or the requested feature already existed. For example, in Issue\#2527, a discussant used animation to show that the ``Multiline Edit'' requested feature already existed.

\subsubsection{Demonstrating an implemented solution}
We found nine cases in which discussants used visual contents to demonstrate a newly implemented solution (i.e., a bug fix or a new feature).

\emph{Showing or testing an implemented feature.} In five cases, visual contents were used for demonstrating working implementations of a new feature, a feature enhancement, or a design change. Discussants used these visual contents either to demonstrate the features that they implemented themselves or to illustrate the results of a feature implemented by another person in the discussant's environment. They may also make one or multiple versions of the feature and ask for feedback. For example, in Issue\#669, the issue reporter used animation to show a working prototype of an enhanced feature for marking cells in order to get community feedback.

\emph{Confirming a fixed problem or bug.} We found four cases in which the issue reporter confirmed, with a visual content, that a problem or a bug is fixed. The fix may or may not involve help from the community and the cause of the bug may or may not be known. For example, in Issue\#4514, the issue reporter discovered a workaround to fix the issue but did not understand why it worked.

\subsubsection{Social purposes.} In one case in our sample, the discussant used a GIF of cats to express happiness and engage in small talk.

\section{Discussion}
In this section, we first discuss the main findings of our analysis. Then we examine the implications of our results to the design of issue tracking systems for facilitating the use of visual contents. 
\subsection{Discussion on the main findings}

Our results show that \textbf{visual contents were commonly used in the issue discussions}. About one-fourth of the issues in our dataset included at least one visual content. Among these issues, about half included more than one visual contents, indicating that the use of visual contents is not accidental. Interestingly, in many cases (781, or 72.9\% among those included a visual content) the issue discussants included a visual content in the original issue post. This is somewhat contradictory to Davies et al.~\cite{Davies2014} and Breu et al.~\cite{Breu2010}, who found that examples and screenshots were often missing in bug reports and had to be requested by the maintainers. This inconsistency may be due to the progressive popularity and inclusiveness of open source projects; in other words, issue posters of Jupyter Notebook (created in 2015) are becoming increasingly aware of the information needs to facilitate an effective discussion.

By looking at the characteristics of issues that included a visual content, we found that issues that only included visual contents in the comments involved longer discussions among a larger number of discussants. 
In other words, \textbf{there is a correlation between (a) the necessity of introducing a visual content in the middle of a discussion and (b) the complexity of the discussion}. However, this may not be a causal relationship. Future investigation is needed to further understand this phenomenon.

Examining the visual contents types, we found that \textbf{visual contents are not limited to communicate user interaction design}. User interfaces only appeared in 63.8\% of the visual contents we examined to communicate information on the main and peripheral UI or highlight certain UI components. In the other cases, visual contents were used to present various types of information, including command-line contents, code snippets, and information in development tools. These results provide evidence against the assumption that visual contents are only relevant to product design.

Our qualitative analysis results on the purpose of using visual contents indicated that \textbf{the issue discussants used visual contents to support discussion on both problem and solution spaces}. In fact, they were used across different problem-solving stages; i.e., to \textit{illustrate the problem} in the very beginning of problem-solving, to \textit{support the investigation of a problem} in the early stages, to \textit{suggest a solution} in potential problem resolution, and to \textit{demonstrate an implemented solution} in the conclusion of problem-solving.

\subsection{Design implications}
Together, our results provided several design implications for issue tracking systems to help facilitate the use of visual contents. Our results indicated diverse types and purposes of visual contents. Additionally, visual contents were posted by various types of contributors, including maintainers, external developers, and end users. Depending on the types and purposes of visual contents, as well as the roles of the community member who posted them, different types of tool support need to be provided. For example, to support the purpose of \textit{suggesting a solution}, collaborative annotation features would be useful in order for the community members to provide their feedback to the proposed solution; for the purpose of \textit{supporting the investigation of a problem}, on the other hand, features that help validate the supportive information provided in the visual contents would further facilitate the discussion. Our results in this work serve as a framework to consider these types of visual content types and purposes when designing the corresponding tools.

Further, during our analysis, we found that visual contents play an essential role in the issue discussions. While visual artifacts themselves need to be accompanied by the text in order to sufficiently support the understanding of the issue discussions, existing work on issue tracking systems seems to be extensively focused on textual analysis. Thus new tools and techniques that leverage information presented in issue tracking systems for supporting software engineering tasks should not overlook the role of visual contents. Incorporating the various textual and visual information would be an essential consideration moving forward. For example, incorporating both textual contents and visual contents into logical augments (e.g., extending the work of~\cite{Wang2020ArguLens}) would enrich the experience of the users of issue tracking systems when examining the various points of view expressed by different discussants. Traceability tools (e.g.,~\cite{Nicholson2020}) can also consider both textual and visual information in order to support a more accurate and enhanced trace-link analysis and construction.





\section{Limitations and threats to Validity}
First, we only focused on one project (i.e., Jupyter Notebook) in our analysis. We chose this project because it has the potential of involving a diverse type of visual contents. However, future work is needed to establish the generalizability of our findings to other projects and communities. Similarly, our analysis is focused on discussions that happened on GitHub Issues. Although GitHub is an increasingly popular platform hosting open source projects, the features provided by the platform may have influenced the characteristics of the visual contents used. Future studies on other platforms are needed to extend our work.

Second, due to the manual effort required for the qualitative analysis, we were only able to analyze 50 threads that included visual contents. While we have achieved saturation in our analysis, there may be other types and purposes of visual contents that we missed in our sample. Further, the coding process during the content analysis may involve the subjectivity of the researchers. We mitigated this risk through a rigorous qualitative methodology that involved multiple researchers and examined inter-rater reliability.

Finally, our analysis is only focused on the explicit contents created by the issue discussants. Our results are thus limited to the behaviors observable on the issue tracking systems. We could not identify, for example, discussants' decision-making process for adding a visual content, or any reasons for not using a visual content. Future work may investigate these topics through user studies such as interviews and surveys.





\section{Conclusion}
In this paper, we aimed at understanding the characteristics of visual contents used in open source software issue discussions. Our analysis on the Jupyter Notebook project indicated that a large quantity of the issue discussions included one or more visual contents, either in the issue posts or the issue comments. Through a qualitative content analysis, we also identified eight general types of visual contents used in those issues, served as diverse purposes across different problem-solving stages during issue discussion. This work serves as an important step towards building more comprehensive knowledge on the use of visual artifacts in open source issue discussions. Our results provided implications and suggestions for both researchers and practitioners who aim at investigating and enriching the issue discussion experience in order to engage diverse open source community members.

\begin{acks}
This work is partially supported by Alfred P. Sloan Foundation (Grant No.: G-2021-16745).
\end{acks}

\bibliographystyle{ACM-Reference-Format}
\bibliography{references}

\begin{figure*}[t]
    \centering
    \includegraphics[width=0.82\textwidth]{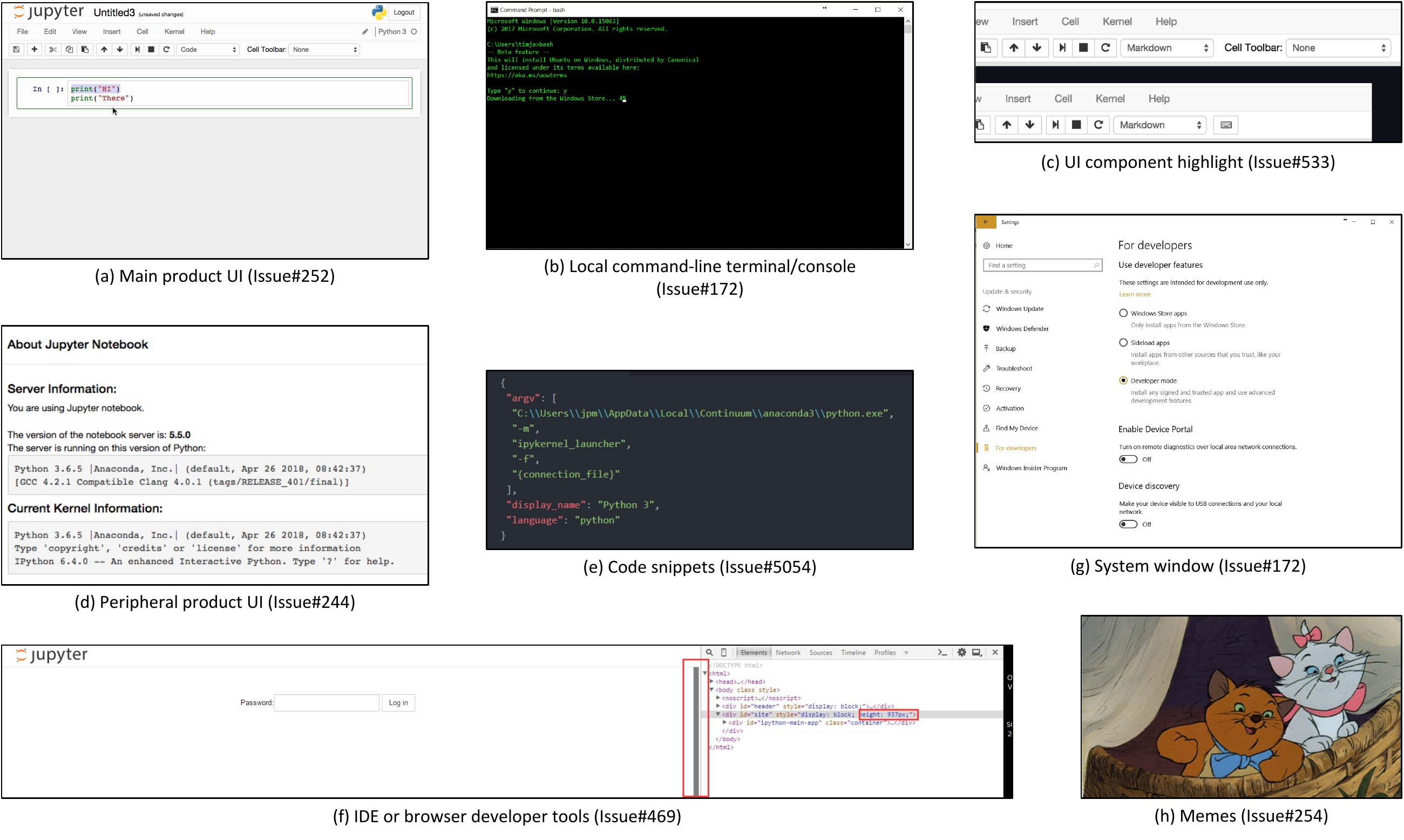}
    \caption{Examples of different types of visual contents}
    \label{fig:content-type-examples}
\end{figure*}

\begin{figure*}[t]
    \centering
    \includegraphics[width=0.82\textwidth]{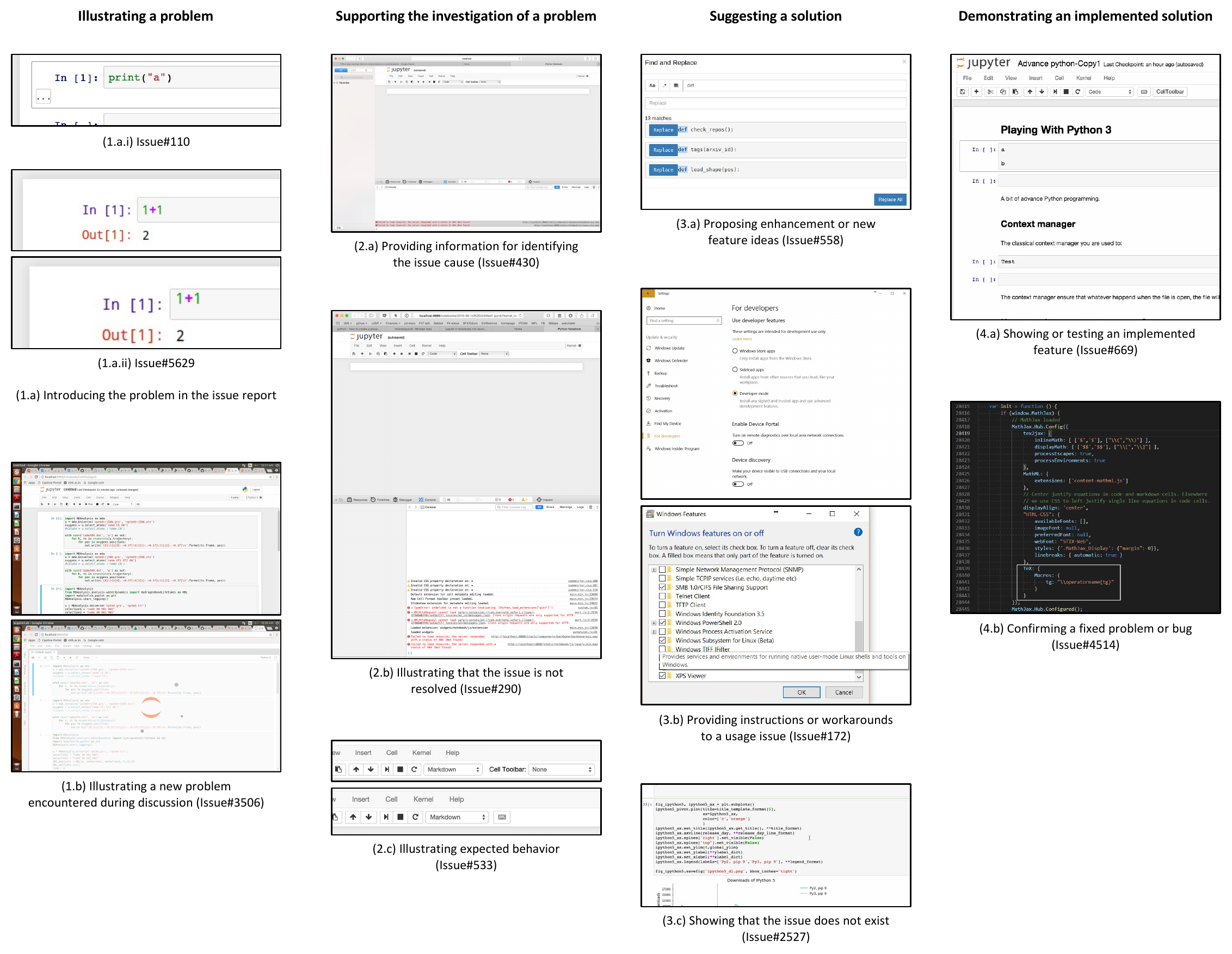}
    \caption{Examples of different types of purpose}
    \label{fig:purpose-type-examples}
\end{figure*}

\end{document}